\newcommand{\useepj}{2}
\ifcase\useepj\or
\documentclass[epj,nopacs
,draft
               ]{svjour}
\usepackage{amsmath,latexsym,epsfig}
\input{epjtitlepage}
\or
\documentclass[12pt]{article}
\usepackage{amsmath,epsfig,a4,times}
\fi

\newcommand{\lambdabar}{{\overline\lambda}}
\newcommand{\sigmabar}{{\overline\sigma}}
\newcommand{\psibar}{{\overline\psi}}
\newcommand{\epsilonbar}{{\overline\epsilon}}
\newcommand{\thetabar}{{\overline\theta}}
\newcommand{\chibar}{{\overline\chi}}

\newcommand{\alphadot}{{\dot\alpha}}
\newcommand{\betadot}{{\dot\beta}}

\newcommand{\intx}{\int d^4x}
\def\dg#1{\frac{\delta\Gamma}{\delta#1}}
\def\dF#1{\frac{\delta{\cal F}}{\delta#1}}
\def\dfunc#1#2{\frac{\delta^{#1}}{\delta#2}}
\def\dfrac#1#2{\frac{\delta{#1}}{\delta{#2}}}
\def\dpartial#1#2{\frac{\partial^{#1}}{\partial#2}}
\def\Number#1{#1\dfunc{}{#1}}
\def\pslash#1{{\setbox0=\hbox{$#1$}
  \rlap{\ifdim\wd0>.7em\kern.22\wd0\else\kern.1\wd0\fi /}#1}}

\def\Sym{{\rm Sym}}

\allowdisplaybreaks[1]
\begin{document}
\ifcase\useepj\or
\maketitle
\or
\begin{titlepage}
\begin{flushright}
KA--TP--16--2000\\
BN--TH--00--07\\
{\tt hep-ph/0007134}\\
\end{flushright}
\vspace{3ex}
\begin{center}
{\Large\bf Renormalization of supersymmetric\\[.2ex] Yang-Mills theories with soft
    supersymmetry\\[.2ex] breaking\\}
\vspace{3ex}
{\large W. Hollik$^a$,
          E. Kraus$^b$,
          D. St{\"o}ckinger$^a${\renewcommand{\thefootnote}{\fnsymbol{footnote}}
\footnote{\parbox[t]{10cm}{
          hollik@particle.physik.uni-karlsruhe.de,\\
          kraus@th.physik.uni-bonn.de,\\
          ds@particle.physik.uni-karlsruhe.de.}}}
  \\[2ex]
  \parbox{10cm}{\small\center\em
           $^a$ Institut f{\"u}r Theoretische Physik, 
                Universit{\"a}t Karlsruhe,
  \\            D--76128 Karlsruhe, Germany
  \\
           $^b$ Physikalisches Institut,
                Universit{\"a}t Bonn,
  \\
               Nu{\ss}allee 12, D--53115 Bonn, Germany}
  }
\end{center}
\vspace{2ex}
\begin{abstract}
The renormalization of supersymmetric Yang-Mills
theories with soft supersymmetry breaking is presented using spurion
fields for introducing the breaking terms. 
It is proven that
renormalization of the fields and parameters in the classical
action yields precisely the correct counterterms to cancel all
divergences. 
In the course of the construction of higher
orders additional independent parameters appear, but they can be
shown to be irrelevant in physics respects. Thus, the only
parameters with influence on physical amplitudes are the
supersymmetric and the well-known soft breaking parameters.
\end{abstract}
\end{titlepage}
\fi

\section{Introduction}

At future experiments at the LHC or at a linear $e^+e^-$
collider, supersymmetric extensions of the standard model can be
tested decisively~\cite{EXP}.
On the theoretical side, exploiting the potential
of these experiments requires a thorough control of the quantization
and the renormalization of supersymmetric models.
One important characteristic of supersymmetric extensions of the
standard model is the appearance of so-called soft
supersymmetry-breaking terms \cite{GG}. Models with soft-breaking
terms have been renormalized using the Wess-Zumino gauge in ref.\
\cite{MPW96b}. The construction in \cite{MPW96b} yields a 
result with an inherent ambiguity.
There appear new kinds of parameters that have no
interpretation as either supersymmetric or soft-breaking parameters. 
Hence, it is unclear 
whether these extra parameters constitute a new kind of free, in principle 
measurable, input parameters, and how the results would influence
the relation to phenomenology.
This effect can be understood as a consequence of the construction using a
BRS doublet for introducing the soft breaking.

In the present article, an alternative approach to the renormalization
of softly broken supersymmetric gauge theories is
presented using the spurion fields introduced originally in
\cite{GG}. Since the spurion fields are  
supermultiplets by themselves, soft breakings of supersymmetry are
distinguished from soft breakings of gauge invariance and other
non-standard breakings (see e.g.\ \cite{JJ}). Since the spurion 
fields are dimensionless, they can appear in arbitrary powers in 
the action --- thus in our approach there appear new parameters, too. 
We can prove, however, that the additional parameters do not influence
physical amplitudes and hence are irrelevant in physics respects.

For the characterization of the symmetries, a Slavnov-Taylor identity 
of the same structure as in the unbroken case \cite{White92a,MPW96a} 
can be used. Since no supersymmetric and gauge invariant regularization
is known, we do not rely on the existence of such a scheme and 
define all Green functions, using the algebraic method,  
via the Slavnov-Taylor identity.
On this basis the relations between the renormalization of soft
and supersymmetric parameters, given in \cite{Yamada,NRTGG,NRTs,SQEDNRT}, 
are not  included in the construction; 
all soft-breaking terms can appear with arbitrary
renormalization constants. As demonstrated for supersymmetric QED 
in \cite{SQEDNRT}, a derivation of such results requires a much more
sophisticated introduction of the soft-breaking terms
and is beyond the pure proof of renormalizability.

We restrict ourselves to a simple, non-Abelian gauge group and exclude
spontaneous symmetry breaking and CP violation. Together with the
treatment of the intricacies of the standard model due to its
spontaneously broken, non-semi\-simple gauge group \cite{Kraus97} and
supersymmetric non-abe\-lian \cite{White92a,MPW96a} and Abelian
\cite{HKS99} gauge theories without soft breaking, this should provide
the necessary building blocks for the renormalization of the
supersymmetric extensions of the standard model.

The outline of the present article is as follows. In
sec.~\ref{Sec:Model} the basic notions of the considered models and of
soft supersymmetry breaking are introduced. 
In sec.~\ref{Sec:Quantization} the
symmetry identities describing gauge invariance and softly broken
supersymmetry are constructed according to the basic idea described
above. 

Sections \ref{Sec:Renormalization}, \ref{Sec:PhysicalPart} constitute
the main part of the paper.
In sec.~\ref{Sec:Renormalization} it is shown that---similar to the
case of \cite{MPW96b}---by introducing 
the external chiral multiplet an infinite number of parameters appears
in the most general classical action. 
That these parameters are all
irrelevant in physics respects and do not even appear in practice is
demonstrated in sec.~\ref{Sec:PhysicalPart}. The theorems proven there
are our central results and finally also imply that all divergences
can be absorbed in accordance with the symmetries.
In sec.~\ref{Sec:Alternatives} our approach is compared to the one of
\cite{MPW96b} 
and its advantages and disadvantages are discussed.
In the appendix our conventions and the BRS transformations are collected.

\section{The model and its symmetries}
\label{Sec:Model}

\subsection{Supersymmetric part}

We consider supersymmetric Yang-Mills theories with a simple gauge
group, coupled to matter. In this class of models there are the
following fields: 
\begin{itemize}
\item One Yang-Mills multiplet
  in the adjoint representation of the gauge group. 
  This multiplet consists of the spin-1 gauge fields $A^\mu_a$ and the
  spin-$\frac12$ gauginos $\lambda^\alpha_a,\lambdabar_a{}_\alphadot$.
\item Chiral supersymmetry multiplets $(\phi_i,\psi_i^\alpha)$ for the
  matter fields consisting of scalar and spin-$\frac12$ fields that
  transform under a representation of the gauge group which is in
  general reducible. The corresponding hermitian generators are called
  $T^a_{ij}$.
\end{itemize}
This minimal set of fields corresponds to the Wess-Zumino
gauge and is used throughout the whole paper. Still it will be convenient to have
the compact superspace notation at hand. In superspace, fermionic
variables $\theta^\alpha$, $\thetabar^\alphadot$ and covariant derivatives
$
D_\alpha  =  \frac{\partial}{\partial\theta^\alpha} -
i(\sigma^\mu\overline\theta)_\alpha\partial_\mu, 
\overline{D}_{\dot\alpha}  = 
\frac{\partial}{\partial\thetabar^{\dot\alpha}} +
i(\theta\sigma^\mu)_{\dot\alpha}\partial_\mu, 
$ are used, and the
fields introduced above are combined in the following vector, chiral
and field strength superfields\footnote{For the vector superfield, the
  Wess-Zumino 
  gauge is used.}
\begin{eqnarray} 
V_a(x, \theta, \overline{\theta})
& = & \theta\sigma^\mu\overline\theta A_a{}_\mu(x)
+i\theta\theta\overline\theta\overline\lambda_a(x) -
i\overline\theta\overline\theta\theta\lambda_a(x)
\nonumber\\
&&{} + \frac{1}{2}\theta\theta\overline\theta\overline\theta D_a(x)\ ,\\
\Phi_i(y,\theta) & = & \phi_i(y) + 
                       \sqrt{2}\ \theta\psi_i(y) + \theta\theta F_i(y) 
\ ,\\
\label{WalphaDef}
W_\alpha & = & -\frac{1}{8g}\overline{D}\overline{D}
(e^{-2gV}D_\alpha e^{2gV})
\end{eqnarray}
with the chiral coordinate $y^\mu = x^\mu -
i\theta\sigma^\mu\overline{\theta}$ and $V=T^a V_a$, $W_\alpha=T^a
W_{a\alpha}$. 
Whenever we use a
superspace expression, it is understood that the auxiliary fields $D_a$
and $F_i$ are eliminated by their respective equations of motion
derived from the complete classical action,
$\frac {\delta\Gamma_{\rm cl}}{\delta{D_a}}
=\frac {\delta\Gamma_{\rm cl}}{\delta{F_i}}
=\frac {\delta\Gamma_{\rm cl}}{\delta{F^\dagger_i}}=0$.

Using this notation and superspace integrals with the normalization
$
\int d^2\theta\ \theta\theta = \int d^2\thetabar\ \thetabar\thetabar =
1,
$
the supersymmetric part of the classical action reads 
\begin{eqnarray}
\Gamma_{\rm susy} & = & 
  \intx \, d^2\theta\, d^2\thetabar\ \Phi^\dagger e^{2gV} \Phi 
\nonumber\\&&{}
+ \Bigl(\intx\, d^2\theta\ \frac{1}{4} W_a^{\alpha}W_a{}_\alpha
        + W(\Phi) + h.c.\Bigr)
\label{GammaSusy}
\end{eqnarray}
with the superpotential\footnote{Gauge singlets are excluded.}
\begin{eqnarray}
W(\Phi) & = & \frac{m_{ij}}{2} \Phi_i\Phi_j +
\frac{g_{ijk}}{3!}\Phi_i\Phi_j\Phi_k \ .
\end{eqnarray}

\subsection{Soft supersymmetry breaking}
\label{Sec:SoftBreaking}

Soft-breaking terms break supersymmetry without destroying its
attractive features. In the present work we restrict the soft-breaking
terms to the terms found and classified by
Girardello and Grisaru (GG)~\cite{GG}. 
Their list of soft-breaking terms is quite short: 
\begin{itemize}
\item mass terms for scalar fields, $-M_{ij}^2\phi_i^\dagger \phi_j$,
\item holomorphic bilinear and trilinear terms in the scalar
  fields,\newline
$-(B_{ij} \phi_i \phi_j + A_{ijk} \phi_i \phi_j \phi_k + h.c.)$,
\item mass terms for gauginos,
$\frac{1}{2} \left( M_\lambda\lambda_a \lambda_a + h.c.\right)$.
\end{itemize}
These GG terms have two crucial properties: First, they break
supersymmetry without introducing quadratic divergences. And
second, they may be viewed as a part of a {\em power-counting
renormalizable and supersymmetric} interaction term with an external
supermultiplet (spurion) \cite{GG}. This can be shown by introducing
one external chiral multiplet with $R$-weight $0$, mass dimension $0$
and a constant shift in its $\hat{f}$ component\footnote{The $\hat{f}$
  component of this external chiral superfield need not be eliminated
  since $\hat{f}$ is no dynamical field and does not satisfy
  particular equations of motion.}: 
\begin{eqnarray}
\eta(y,\theta) & = & a(y) + \sqrt{2} \theta\chi(y)
                     + \theta\theta \hat{f}(y),\\
\hat{f}(y)     & = & f(y)+f_0.
\end{eqnarray}
Then the supersymmetric extensions of the above soft breaking terms
can easily be written in superspace:
\begin{eqnarray}
\Gamma_{\rm soft}
& = & -  \intx \, d^2\theta\, d^2\thetabar\ 
       \tilde{M}_{ij}^2 \eta^\dagger\eta \Phi_i^\dagger (e^{2gV}\Phi)_j 
\nonumber\\
&&{}  -  \intx \, d^2\theta\ (\tilde{B}_{ij} \eta\Phi_i\Phi_j
                     +\tilde{A}_{ijk}\eta\Phi_i\Phi_j\Phi_k) + h.c.
\nonumber\\
&&{}  -  \intx \, d^2\theta\ \frac12\tilde{M}_\lambda
                      \eta W^\alpha_a W_a{}_\alpha + h.c.
\label{GGSoftTerms}
\end{eqnarray}
As long as $\eta$ and its component fields are treated as external
fields with arbitrary values, these interaction terms are
manifestly supersymmetric. Only in the limit
\begin{eqnarray}
a(x)            =
\chi(x)         =
f(x)            & = & 0,\nonumber\\
\eta(x,\theta)  & = & \theta\theta f_0,
\label{PhysicalLimit}
\end{eqnarray}
they reduce to the soft breaking terms with $\tilde{M}_{ij}^2 |f_0|^2
 = M_{ij}^2$, $\tilde{B}_{ij} f_0 = {B}_{ij}$,
 $\tilde{A}_{ijk}f_0=A_{ijk}$, $\tilde{M}_\lambda f_0=M_\lambda$.

The GG soft breaking terms comprise all possible terms of mass
dimension 2 but not all possible terms of mass dimension 3. Obviously,
not only $\lambda\lambda$ and $\phi\phi\phi$ but also $\psi\psi$ and
$\phi^\dagger\phi\phi$ are supersymmetry-breaking terms of mass
dimension 3.\footnote{For instance, in the case of the minimal supersymmetric
standard model the $\phi\phi\phi$ GG terms are (we adopt the
conventions of ref.\ \cite{JJ})
$$
m_{10}\lambda_t H_2 Q \bar{t} + m_{8}\lambda_b H_1 Q \bar{b}
+ m_{6}\lambda_\tau H_1 L \bar{\tau}\ ,
$$
whereas the following non-GG terms are also perfectly gauge-invariant
supersymmetry-breaking terms that do not induce quadratic divergences:
$$
m_9\lambda_t H_1^* Q \bar{t} + m_{7}\lambda_b H_2^* Q \bar{b}
+ m_{5}\lambda_\tau H_2^* L \bar{\tau}\ .
$$
}
The terms of the form $\psi\psi$ and $\phi^\dagger\phi\phi$ are
excluded from the GG class 
because in
general they introduce quadratic divergences. However, as mentioned
e.g.\ in \cite{JJ}, in many concrete models, like the minimal
supersymmetric extension of the standard model, these quadratic
divergences are absent. Therefore, concerning only the quadratic
divergences, the GG class is too narrow. 

If soft breaking is introduced via the coupling to $\eta$, the non-GG
terms are excluded, since they cannot be extended to a power-counting
renormalizable and supersymmetric interaction such as in
(\ref{GGSoftTerms}). This means that the possible supersymmetric
coupling to the spurion $\eta$ is the more profound characterization
of the GG soft breaking terms than absence of quadratic divergences.


\section{Quantization}
\label{Sec:Quantization}
\subsection{Construction of the Slavnov-Taylor identity}

Our aim is now to find a definition of supersymmetric gauge theories
with soft breaking. Analogously to the case without soft breaking,
softly broken supersymmetry should be combined with gauge invariance
in a single Slavnov-Taylor identity. Since soft breaking terms are
characterized by the possible coupling to the external $\eta$
multiplet, there is the following possibility: The Slavnov-Taylor
identity has the same form as in the unbroken case but it contains
also the $\eta$ multiplet. In this way, first a fully supersymmetric
model is described. Then $\eta$ is set to the constant
(\ref{PhysicalLimit}), and in this way the soft breaking is
introduced.

According to this approach, the Slavnov-Taylor identity is constructed
along the same lines as in the unbroken case \cite{MPW96a}. The basic
elements of the construction are the following: First, 
BRS transformations are introduced at the classical level. Since
supersymmetry, gauge transformations, and translations are deeply
entangled in the Wess-Zumino gauge, all three symmetries have to be
combined into the BRS transformations $s$, and three kinds of ghost
fields have to be used. These are the 
fields 
\begin{eqnarray}
c_a(x), \epsilon^\alpha,\epsilonbar^\alphadot,\omega^\nu,
\end{eqnarray}
corresponding to gauge and supersymmetry transformations
and translations, respectively. Only the Faddeev-Popov ghosts $c_a$ are 
quantum fields, whereas the other ghosts are space-time independent
constants because the corresponding symmetries are global. The
statistics of all ghost fields is opposite to the one required by the
spin-statistics theorem. The
explicit form of the BRS transformations can be found in the
appendix.

Second, the sum of the gauge fixing and ghost terms has to be BRS
invariant in 
order to ensure the decoupling of the unphysical degrees of freedom
and the unitarity of the physical S-matrix. Thus it can be obtained as
the BRS transformation of some fermionic expression with ghost number
$-1$. In order to define such an expression we introduce the antighosts
$\bar{c}_a(x)$ and auxiliary fields $B_a$ 
and write the usual renormalizable gauge fixing term with
arbitrary gauge parameter $\xi$ and a linear gauge fixing function
$f_a = \partial_\mu A^\mu_a$ as
\begin{eqnarray}
\Gamma_{\rm fix,\ gh} & = & \intx\ s[\bar{c}_a (f_a + \frac{\xi}{2}B_a)]
\ .
\end{eqnarray}

Third, most of the BRS transformations are non-linear in
the propagating fields and thus affected by quantum corrections. In
order to cope with the renormalization of the composite operators
$s\varphi_i$ we couple them to external fields $Y_i$:
\begin{eqnarray}
\Gamma_{\rm ext} & = & \intx \Bigl(Y_{A^\mu_a} sA^\mu_a
+ Y_{\lambda_a}^\alpha s\lambda_{a\alpha}
+ Y_{\lambdabar_a}{}_\alphadot s\lambdabar_a^\alphadot
\nonumber\\&&{}\hfill
+ Y_{\phi_i} s\phi_i + Y_{\phi_i^\dagger} s\phi_i^\dagger
+ Y_{\psi_i}^\alpha s\psi_i{}_\alpha
+ Y_{\psibar_i}{}_\alphadot s\psibar_i^\alphadot
\nonumber\\&&{}\hfill
+ Y_{c_a} sc_a\Bigr) \ .
\label{GammaExt}
\end{eqnarray}
Note that the implicit elimination of the $D_a$ and $F_i,
F_i^\dagger$ fields yields additional bilinear terms in the external
$Y$ fields. 
Using the external $Y$ fields we can write down the Slavnov-Taylor
operator $S(\cdot)$ corresponding to the BRS operator $s$. Acting on a
general functional ${\cal F}$ it reads:
\begin{eqnarray}
S({\cal F}) & = & S_0({\cal F}) + S_{\rm soft}({\cal F})\ ,
\label{STOperator}
\\
S_0({\cal F}) & = &
\intx\Bigl(\dF{Y_{A^\mu_a}}  \dF{A_a^\mu}
+ \dF{Y_{\lambda_a}{}_\alpha}\dF{\lambda_a^\alpha}
+ \dF{Y_{\lambdabar_a}^\alphadot}\dF{\lambdabar_a{}_\alphadot}
\nonumber\\&&{}\quad
+ \dF{Y_{\phi_i}}\dF{\phi_i}
+ \dF{Y_{\phi_i^\dagger}}\dF{\phi_i^\dagger}
+ \dF{Y_{\psi_i}{}_\alpha}\dF{\psi_i^\alpha}
\nonumber\\&&{}\quad
+ \dF{Y_{\psibar_i}^\alphadot}\dF{\psibar_i{}_\alphadot}
\nonumber\\&&{}\quad
+ \dF{Y_{c_a}}\dF{c_a} + s\bar{c_a} \dF{\bar{c_a}}
+ s B_a \dF{B_a}\Bigr)
\nonumber\\&&{}
+ s\epsilon^\alpha\frac{\partial{\cal F}}{\partial\epsilon^\alpha}
+ s\epsilonbar_\alphadot
  \frac{\partial{\cal F}}{\partial\epsilonbar_\alphadot}
+ s\omega^\nu \frac{\partial{\cal F}}{\partial\omega^\nu}
\label{SusySTOp}
\ ,\\
S_{\rm soft}({\cal F}) & = &
\intx\Bigl(sa\dF{a} + sa^\dagger\dF{a^\dagger}
+ s\chi^\alpha\dF{\chi^\alpha} 
\nonumber\\&&{}\quad
+ s\chibar_\alphadot\dF{\chibar_\alphadot}
+ sf\dF{f} + sf^\dagger\dF{f^\dagger}
\Bigr)\ .
\end{eqnarray}
Only the linear BRS transformations appear explicitly here.

\subsection{Defining symmetry identities}

Now we are in the position to spell out the complete definition of the
symmetries of the model as a set of requirements on the effective
action $\Gamma$, the quantum extension of the classical action
$\Gamma_{\rm cl}$ and the generating functional of one-particle
irreducible vertex functions:
\begin{itemize}
\item Slavnov-Taylor identity:
\begin{eqnarray}
S(\Gamma) & = & 0\ .
\label{STI}
\end{eqnarray}
\item Gauge fixing condition:
\begin{eqnarray}
\dg{B_a} & = & \frac{\delta\Gamma_{\rm fix}}{\delta B_a}
 = f_a + \xi B_a \ .
\label{GaugeFixing}
\end{eqnarray}
\item Translational ghost equation:
\begin{eqnarray}
\dg{\omega^\nu} & = & \frac{\delta\Gamma_{\rm ext}}{\delta\omega^\nu}
=\intx\sum_{\varphi_i}(-1)^{GP_i}Y_i
i\partial_\nu\varphi_i
\label{OmegaEq}
\end{eqnarray}
with $\Gamma_{\rm ext}$ in eq.\ (\ref{GammaExt}), and where
$(\varphi_i,Y_i)$ runs over the dynamical fields with corresponding
$Y$ fields and $GP_i$ denotes the Grassmann-parity of $\varphi_i$.
\item Global symmetries: We require $\Gamma$ to be invariant under CP
  conjugation and under global gauge transformations and continuous
  $R$-trans\-for\-mations and to preserve ghost number (see table
  \ref{Tab:QuantumNumbers}). There may be
  further symmetries such as lepton number conservation, but these we
  leave unspecified. We only assume that the global symmetries exclude
  mixings between the $\psi_i$ and the $\lambda_a$, between $\phi_i$
  and $\phi_j^\dagger$  and between the combinations $\hat{f}\phi_i$
  and $(\hat{f}\phi_j)^\dagger$.
\begin{table}[tbh]
\begin{displaymath}
\begin{array}{|c||c|c|c|c|c|c|c|c|c|c|c|c|c|c|c|c|c|c|}
\hline
\chi & A_a^\mu  & \lambda_a^\alpha &  
\phi_i &
\psi^\alpha_i & a & \chi^\alpha & \hat{f} &  
 c_a & \epsilon^\alpha & 
\omega^\nu & \bar{c}_a & B_a 
\\ \hline
R   & 0  & 1 & n_i&n_i-1& 0 & -1 & -2 & 0 & 1  & 0 & 0 & 0 \\ \hline 
Q_c & 0  & 0 &  0 &  0  & 0 & 0  &  0 &+1 & +1 &+1 &-1 & 0 \\ \hline
GP  & 0  & 1 &  0 &  1  & 0 & 1  &  0 & 1 & 0  & 1 & 1 & 0 \\ \hline
dim & 1  &3/2&  1 & 3/2 & 0 & 1/2&  1 & 0 &-1/2&-1 & 2 & 2 \\ \hline
\end{array}
\end{displaymath}
\caption{Quantum numbers. $R,Q_c,GP,dim$ denote $R$-weight and ghost
  charge, Grassmann parity and the mass dimension, respectively. The
  $R$-weights $n_i$ of the chiral multiplets are left arbitrary. The
  quantum numbers of the external fields $Y_i$ introduced in
  sec.~\ref{Sec:Quantization} can be obtained from the
  requirement that $\Gamma_{\rm ext}$ is neutral, bosonic and has
  $dim=4$. The commutation rule for two general fields is
  $\chi_1\chi_2 = (-1)^{GP_1 GP_2} \chi_2\chi_1$.}
\label{Tab:QuantumNumbers} 
\end{table}
\item Physical part: As already stated in sec.~\ref{Sec:SoftBreaking},
  the physical part of the effective action is defined to be 
\begin{eqnarray}
\Gamma|_{a=\chi=f=0}\ .
\end{eqnarray}
  In this limit, already defined in eq.\ (\ref{PhysicalLimit}),
  supersymmetry is softly broken by GG terms.
\end{itemize}
For later use we introduce the abbreviation $\Sym(\Gamma)=0$ for this
set of symmetry requirements:
\begin{eqnarray}
\Sym(\Gamma)=0&\Leftrightarrow&
(\ref{STI}), (\ref{GaugeFixing}), (\ref{OmegaEq}), 
\mbox{Global symmetries.}\quad
\end{eqnarray}

The canonically normalized classical action is given by the sum 
\begin{eqnarray}
\Gamma_{\rm cl,\ canonical} & = & \Gamma_{\rm susy} + \Gamma_{\rm soft} +
\Gamma_{\rm fix,\ gh} + \Gamma_{\rm ext}\ ,
\label{GammaClSpecial}
\end{eqnarray}
with eliminated $D_a$ and $F_i$ fields. The construction guarantees
that $\Sym(\Gamma_{\rm cl,\ canonical})=0$. Its explicit form reads
\begin{eqnarray}
\lefteqn{\Gamma_{\rm cl,\ canonical}|_{a=\chi=0}}\qquad
\nonumber\\
 & = & \Gamma^0_{\rm susy}
  + \Gamma^0_{\rm soft} + \Gamma^0_{\rm fix,\ gh} + \Gamma^0_{\rm ext}
  + \Gamma^0_{\rm bil}\ ,
\label{GammaClGenSoft}
\\
\Gamma^0_{\rm susy} & = & \intx \Bigl(-\frac{1}{4}(F_{\mu\nu}^a)^2
\nonumber\\
&&{} + \frac{i}{2}\overline\lambda^a\overline\sigma^\mu
       (D_\mu\lambda)^a
     + \frac{i}{2}\lambda^a\sigma^\mu 
       (D_\mu\overline\lambda)^a
\nonumber\\
&&{} + (D^\mu \phi)^\dagger(D_\mu \phi)
     + \overline\psi\overline\sigma^\mu iD_\mu\psi
\nonumber\\
&&{} - \sqrt{2} g (i\overline\psi\overline\lambda \phi - i\phi^\dagger
       \lambda\psi)
\nonumber\\
&&{} - \left(\frac{1}{2}\psi_i\psi_j
          \frac{\partial^2 W(\phi)}{\partial \phi_i  \partial \phi_j}
           + h.c. \right)
\nonumber\\
&&{} - \frac{1}{2}(\phi^\dagger gT^a \phi)^2
     - \left|\frac{\partial W(\phi)}{\partial \phi_i}\right|^2
\Bigr)\ ,
\\
\Gamma^0_{\rm soft} & = & \intx\Bigl(
-\tilde{M}_{ij}^2 \hat{f}^\dagger\hat{f} \phi_i^\dagger \phi_j
\nonumber\\&&{}
-\left(\tilde{B}_{ij} \hat{f} \phi_i \phi_j 
  + \tilde{A}_{ijk}\hat{f} \phi_i \phi_j \phi_k + h.c.\right)
\nonumber\\&&{}
+\frac{1}{2} \left( \tilde{M}_\lambda \hat{f}\lambda^a \lambda^a +
h.c.\right)
\Bigr)\ ,
\\
\Gamma^0_{\rm fix,\ gh} & = &
 \intx \Bigl(B_a f_a + \frac{\xi}{2} B_a^2\Bigr) 
      +\Gamma^0_{\rm gh}
\ ,
\\
\Gamma^0_{\rm gh}&=&
\intx\Bigl(- \bar{c}_a\partial_\mu (D^\mu c)_a
\nonumber\\
&&{}
- \bar{c}_a\partial^\mu(i\epsilon\sigma_\mu\lambdabar_a 
                     -i\lambda_a\sigma_\mu\epsilonbar) 
+ \xi i \epsilon\sigma^\nu\epsilonbar
  (\partial_\nu\bar{c}_a)\bar{c}_a \Bigr)
\label{GaugeFixingTerm}
\ ,\\
\Gamma^0_{\rm ext} & = & \Gamma_{\rm ext}|
^{D_a\to -g\phi^\dagger T_a \phi}
_{F_i\to -(\partial W(\phi)/\partial\phi_i)^\dagger}
\ ,
\\
\Gamma^0_{\rm bil} & = & \intx\Bigl(
\frac12(Y_{\lambda_a}  \epsilon + Y_{\lambdabar_a} \epsilonbar)^2 
+ 2(Y_{\psi_i}\epsilon)(Y_{\psibar_i}\epsilonbar)\Bigr) \ .\quad
\end{eqnarray}
As indicated by the superscript $^0$, the part containing the external
$a$ and $\chi$ fields is suppressed here because its concrete form is
not relevant for our discussion, and only the $\hat{f}$ component of
the $\eta$ multiplet is retained. 
Furthermore, we have introduced the gauge covariant derivative 
\begin{eqnarray}
D_\mu & = & \partial_\mu + igT^a A_a^\mu,
\end{eqnarray}
where in the adjoint representation $T^a$ has to be replaced by
$-if^{abc}$ defined by $[T^a,T^b]=if^{abc}T^c$, and the  field
strength tensor 
\begin{eqnarray}
igT^a F_a^{\mu\nu} & = & [D^\mu,D^\nu]\ ,\\
F_a^{\mu\nu} & = & \partial^\mu A_a^\nu-\partial^\nu A_a^\mu
                   - g f^{abc} A^\mu_b A^\nu_c\ .
\end{eqnarray}
More general classical solutions of the symmetry requirements will be
given sec.~\ref{Sec:GenClassicalSolution}.

\section{Renormalization I: Basics}
\label{Sec:Renormalization}

The symmetry identities constitute a rigorous definition of the
considered models. However, it remains to be checked whether the
models defined in this way are renormalizable. In the present section
the usual analysis of the structure of the symmetric counterterms is
applied, and the existence of  infinitely many different types of
symmetric counterterms is found. The role of these counterterms will be
discussed in section \ref{Sec:PhysicalPart}.

\subsection{Generalized classical solution}
\label{Sec:GenClassicalSolution}

In this subsection we assume that the symmetry identities can be
established at each order by adding appropriate counterterms. Once the
symmetries hold at the order $\hbar^n$, there still may arise
divergences and counterterms may be added. Both the divergences and
the counterterms cannot interfere with the symmetries, which means
that both are of the form $\Gamma_{\rm sym}$ with 
\begin{eqnarray}
&&\Sym(\Gamma_{\le\mbox{\scriptsize n-Loop,
    regularized}}+\hbar^n\Gamma_{\rm sym})
\nonumber\\
 & = & \Sym(\Gamma_{\le\mbox{\scriptsize n-Loop,
    regularized}}) + {\cal O}(\hbar^{n+1})\ ,
\end{eqnarray}
which reduces to
\begin{eqnarray}
\Sym(\Gamma_{\rm cl}+\zeta\Gamma_{\rm sym}) & = & {\cal O}(\zeta^2)\ ,
\label{ClSym}
\end{eqnarray}
with some arbitrary infinitesimal parameter $\zeta$, since all
symmetry identities are linear or bilinear in $\Gamma$. $\Gamma_{\rm
  cl}$ is the classical action, i.e. $\Gamma=\Gamma_{\rm cl}+{\cal
O}(\hbar)$.

A model is renormalizable if all divergences can be absorbed by
counterterms corresponding to renormalization of the fields and
parameters in the classical action and if the number of physical
parameters is finite. 

Eq.\ (\ref{ClSym}) shows how to find the general structure of the
possible divergences and counterterms. Since the perturbed action
$\Gamma_{\rm cl}+\zeta\Gamma_{\rm sym}$ is a solution of the
symmetry identities in terms of a local power-counting renormalizable
functional (classical solution), simply the most general of these
classical solutions has to be calculated. 

In this subsection we determine a certain set of classical solutions
with a result reminding of the result of \cite{MPW96b}. Beyond the
supersymmetric and soft breaking parameters there appear new kinds of
free parameters. In fact, our solutions depend on infinitely many
parameters!

One way to obtain classical solutions different from $\Gamma_{\rm cl,\
  canonical}$ in eq.\ (\ref{GammaClSpecial}) is obvious. Since $\eta$
is neutral with respect to all quantum numbers and has dimension 0 it
can appear without any restrictions in the classical action. Indeed, 
\begin{eqnarray}
\Gamma_{\rm susy} + \Gamma_{\rm soft}
& = & 
  \intx \, d^2\theta\, d^2\thetabar\ 
r_{1ij}(\eta,\eta^\dagger)\Phi^\dagger_i( e^{2gV} \Phi)_j
\nonumber\\&&{}
+ \intx\, d^2\theta\ \Bigl(r_2(\eta)W_a^{\alpha}W_{a\alpha}
\nonumber\\&&{}\quad
      - r_{3ij}(\eta)\Phi_i\Phi_j
      - r_{4ijk}(\eta)\Phi_i\Phi_j\Phi_k\Bigr)
+ h.c.
\end{eqnarray}
is a  possible generalization of (\ref{GammaSusy}), (\ref{GGSoftTerms})
that maintains all symmetry properties of $\Gamma_{\rm cl}$. Here
$r_1$ is an arbitrary real function of $\eta,\eta^\dagger$, and
$r_2,r_3,r_4$ are holomorphic functions of $\eta$. Expanding
$r_1\ldots r_4$ in a Taylor series leads to infinitely
many interaction terms in $\Gamma_{\rm cl}$. The fact that this
generalized action is still symmetric means that to all of these terms
there can be divergent loop contributions and that to each of them a
normalization condition is needed.

There is a further, more complicated way to perturb a classical
solution of the symmetry requirements. We can modify the superfields
appearing in $\Gamma_{\rm susy}$ 
and $\Gamma_{\rm soft}$ by terms depending on $a,\chi,f$. If these
modifications are accompanied by suitable changes in the BRS
transformations in $\Gamma_{\rm ext}$, again classical solutions
are obtained. One specific possibility is the following 
modification of the chiral superfields:
\begin{eqnarray}
\Phi_i & = & u_{1ij}(a,a^\dagger)\phi_j +
 \sqrt{2}(u_1u_2)_{ij}(a,a^\dagger)
 \theta\psi_j
\nonumber\\&&{}
            - \sqrt{2}(u_1u_3)_{ij}(a,a^\dagger)\theta\chi\phi_j 
            + \theta\theta F_i
\label{NewPhi}
\ ,
\end{eqnarray}
where this modification is parametrized by three
arbitrary functions $u_{1},u_2,u_3$ of $a$ and $a^\dagger$. These
fields $\Phi_i$ transform as chiral superfields if the BRS
transformations and thus $\Gamma_{\rm ext}$ is redefined as
\begin{eqnarray}
\lefteqn{\Gamma_{\rm ext}^{\rm \phi,\psi-Part}  =  \intx\Bigl(
Y_{\phi_i}
\Bigl[\sqrt2 u_{2ij}\epsilon\psi_j
 - (u_1^{-1}s_\epsilon  u_{1})_{ij}\phi_j}\quad
\nonumber\\&&{}
     -\sqrt2 u_{3ij}\epsilon\chi\phi_j\Bigr]
\nonumber\\&&{}
- Y_{\psi_i}{}_\alpha
      \Bigl[
       -(u_2^{-1}u_1^{-1}s_\epsilon u_1 u_2)_{ij}\psi_j^\alpha
\nonumber\\&&{}
       +\sqrt2 (u_2^{-1}u_3 u_2)_{ij} \epsilon\psi_j \chi^\alpha
       -\sqrt2 (u_2^{-1}u_3u_3)_{ij}\epsilon\chi\phi_j\chi^\alpha
\nonumber\\&&{}
       +(u_2^{-1}u_1^{-1}(s_\epsilon u_1u_3) 
                -u_2^{-1}u_3u_1^{-1}(s_\epsilon u_1))_{ij}\phi_j\chi^\alpha
\nonumber\\&&{}
       -\sqrt2 i (\epsilonbar\sigmabar^\mu)^\alpha u_{2ij}^{-1}\big(D_\mu\phi_j
\nonumber\\&&{}\qquad
           + (u_1^{-1}\partial_\mu u_1)_{jk}\phi_k + u_{3jk}\phi_k\partial_\mu a\big)
\nonumber\\&&{}
       +\sqrt2\epsilon^\alpha (u_1 u_2)^{-1}_{ij} F_j 
       +\sqrt2\epsilon^\alpha (u_2^{-1}u_3)_{ij} \phi_j \hat{f}
      \Bigr]
\nonumber\\&&{}
+h.c.
+\mbox{ Terms involving } c,\omega^\nu \Bigr)\ .
\label{ModGammaExtPhiPsi}
\end{eqnarray}
Here $s_\epsilon$ denotes only the $\epsilon,\epsilonbar$-dependent
part of the BRS transformation. The terms involving $c,\omega^\nu$ are
identical to those in (\ref{GammaExt}). 
Using $\Phi_i$ from (\ref{NewPhi}) in $\Gamma_{\rm susy,\ soft}$
together with the redefined $\Gamma_{\rm ext}$, we obtain a further
set of classical solutions.

Analogously, the vector superfield and the corresponding part of
$\Gamma_{\rm ext}$ can be modified as follows:
\begin{eqnarray}
V & = & v_1(a,a^\dagger)(\theta\sigma^\mu\thetabar A_\mu 
\nonumber\\&&{}
       +i\theta\theta\thetabar(\lambdabar v_2(a,a^\dagger) + \sigmabar^\mu\chi
       A_\mu v_3(a,a^\dagger))
\nonumber\\&&{}
       -i\thetabar\thetabar\theta(\lambda v_2(a,a^\dagger) - \sigma^\mu\chibar
       A_\mu v_3(a,a^\dagger)))
\nonumber\\&&{}
 + \frac12\theta\theta\thetabar\thetabar D
\ ,
\\
\lefteqn{\Gamma_{\rm ext}^{A_\mu,\lambda-\rm Part}
 =  \intx\Bigl(Y_{A_{a\mu}} \Bigl[
  i\epsilon\sigma_\mu(\lambdabar v_2 + \sigmabar^\nu\chi A_{a\nu} v_3)}\quad
\nonumber\\&&{}
- i(\lambda_a v_2 + \chibar\sigmabar^\nu A_{a\nu} v_3)\sigma_\mu\epsilonbar
- A_{a\mu} (v_1^{-1}s_\epsilon v_1)\Bigr]
\nonumber\\&&{}
+\Bigl(- Y_{\lambda_a}{}_\alpha\Bigl[
\frac{i}{2}\epsilon\sigma^{\rho\sigma}(v_1v_2)^{-1}
F_{a\rho\sigma}(v_1 A)
\nonumber\\&&{}
 + i(v_1v_2)^{-1} \epsilon D_a 
+ \sqrt2 v_3 v_2^{-1} \hat{f}^\dagger \epsilonbar\sigmabar^\mu
A_{a\mu}
\nonumber\\&&{}
-(v_1^{-1}v_2^{-1}s_\epsilon v_1v_2) \lambda_a
-\big[i\epsilon\sigma_\mu(\lambdabar_a+v_3v_2^{-1}\sigmabar^\nu\chi A_{a\nu})
\nonumber\\&&{}
 -i(\lambda+v_3v_2^{-1}\chibar\sigmabar^\nu
 A_{a\nu})\sigma_\mu\epsilonbar\big]
  \chibar\sigmabar^\mu v_3
\nonumber\\&&{}
- v_3 v_2^{-1} \sqrt2 i\epsilon\sigma^\nu(\partial_\nu a^\dagger)
         \sigmabar^\mu A_{a\mu}
\nonumber\\&&{}
- (s_\epsilon v_3)v_2^{-1} \chibar\sigmabar^\mu A_{a\mu}\Bigr]
+h.c.\Bigr)
\nonumber\\&&{}
+ \mbox{ Terms involving } c,\omega^\nu\Bigr)\ .
\end{eqnarray}
Here a modified field strength tensor $F_{a\rho\sigma}(v_1A) =
\partial_\rho (v_1A_{a\sigma}) - \partial_\sigma (v_1A_{a\rho}) -
gf^{abc}v_1^2A_{b\rho}A_{c\sigma}$ has been introduced. 

Note that the functions $u_1$, $u_2$, $v_1$, $v_2$ are
$a,a^\dagger$-dependent generalizations of field renormalizations of
the matter and gauge fields. On the other hand, $u_3, v_3$ are new
kinds of parameters corresponding to field renormalizations of the
form 
\begin{eqnarray}
\psi & \to & \psi - u_3\chi\phi\ ,\\
\lambda_\alpha & \to & \lambda_\alpha - v_3(\sigma^\mu\chibar)_\alpha
A_\mu\ .
\end{eqnarray} 

In addition to these modifications,
obviously a field renormalization of the Faddeev-Popov ghost
\begin{eqnarray}
c\to\sqrt{Z_c} c \ ,&&Y_c\to\sqrt{Z_c}^{-1} Y_c
\end{eqnarray}
and renormalization of all parameters appearing in $\Gamma_{\rm cl}$
in eq.~(\ref{GammaClSpecial}) is possible without violating the symmetry
identities. 

We conclude that the supersymmetry algebra is unstable in the sense
that it allows for arbitrary functions $u_{1,2,3}$ and $v_{1,2,3}$
with an infinite number of Taylor coefficients that have to be
renormalized. So, even without calculating the classical solution to the
symmetry identities in full generality, we know that infinitely many
normalization conditions are needed and the effective action $\Gamma$
depends on infinitely many parameters.

In the physical limit $a=\chi=f=0$ or already in the limit $a=\chi=0$,
the functions $r_i$, $u_i$, $v_i$ reduce to usual field
renormalizations and two additional parameters $u_3(0)$,
$v_3(0)$. Taking these two parameters into account, the canonically
normalized classical action $\Gamma_{\rm cl,\ canonical}$ in eq.\
(\ref{GammaClGenSoft}) changes as follows:
\begin{eqnarray}
\lefteqn{\Gamma_{\rm cl,\ canonical}|_{a=\chi=0}  }
\nonumber\\
& = &
 \Gamma_{\rm cl,\
  canonical}^{\rm eq.\ (\ref{GammaClGenSoft})}|_{a=\chi=0}
\label{GammaClGen}
\nonumber\\
&&{}
+\intx \Bigl(-Y_{\psi_i}{}_\alpha
       (\sqrt2\epsilon^\alpha \hat{f}u_{3ij}(0) \phi_j)
\nonumber\\&&{}
- Y_{\lambda_a}{}_\alpha \sqrt2 v_3(0)\hat{f}^\dagger
\epsilonbar\sigmabar^\mu{}^\alpha A_{a\mu}+h.c.\Bigr)
\ .
\end{eqnarray}
Only the external field part is influenced by the new parameters.

\subsection{Remarks on anomalies}

In the preceding subsection we have assumed that the symmetry
identities can be maintained at each order of perturbation theory.
In principle this need not be true, because there could be
anomalies. For unbroken supersymmetric Yang-Mills theories it is well
known that the only possible anomaly is the supersymmetric extension
of the chiral gauge anomaly \cite{PiSi84,White92a,MPW96a}. In
particular, the relevant cohomology does not depend on the chiral
multiplets at all. In spite of the soft breaking, the formulation of
our model is the same as the one for unbroken supersymmetric
Yang-Mills theories except for the appearance of the additional
chiral $\eta$ multiplet of dimension 0. Therefore, we assume that our
model is anomaly free and the symmetry identities can be restored by
suitable counterterms at each order. 

However, one also has to check for infrared anomalies, i.e. breakings
of the symmetry identities that can only be absorbed by counterterms
of infrared dimension less than 4. Using the assignments from
\cite{MPW96b}\footnote{For the spurion field components we use ${\rm
    dim}_{\rm IR}(a)=2$, ${\rm dim}_{\rm IR}(\chi,f)=1$.}, in
principle counterterms of infrared dimension 
$\ge2.5$ could show up. However, there are no such counterterms of
infrared dimension $<4$ that involve at least two propagating
fields. The other ones cannot be inserted in higher order loop
diagrams and thus are harmless, so there are no infrared anomalies. 

\section{Renormalization II: Physical part of the model}
\label{Sec:PhysicalPart}

In general, a model depending on an infinite number of parameters has
no predictive power. But this is not necessarily the case here,
because all physical amplitudes have to be derived from the effective
action $\Gamma$ in the limit (\ref{PhysicalLimit}), $a=\chi=f=0$. And
we have not yet checked which of the parameters can have any influence
on $\Gamma$ in this limit. 

In this section we prove two theorems showing that the infinitely many
unwanted parameters are irrelevant for physical quantities and do not
appear in practical calculations. Thus the number of physical
parameters is finite and the considered models are renormalizable.
And moreover, the set of physical parameters can be identified with
the supersymmetric and soft breaking parameters.


The essentials of the two theorems are the following:
\begin{enumerate}
\item The only quantities $\Gamma|_{a=\chi=Y_i=0}$, i.e.\ Green functions
  without external $a,\chi$ or $Y_i$ fields, depend on are
\begin{itemize}
\item the field renormalization constants $Z_A$, $Z_\lambda$, $Z_c$,
  $Z_\phi$, $Z_\psi$,
\item the gauge coupling $g$,
\item the parameters in the superpotential $m_{ij}, g_{ijk}$,
\item the soft breaking parameters $\tilde{M}^2_{ij}, \tilde{B}_{ij},
  \tilde{A}_{ijk}, \tilde{M}_\lambda$.
\end{itemize}
  More details and the proof can be
  found in subsec.\ \ref{Sec:PhysParameters}.
\item 
  In practical calculations it is sufficient to solve the symmetry
  identities in the limit $a=\chi=0$,
\begin{eqnarray}
\Sym(\Gamma)|_{a=\chi=0}=0\ .
\end{eqnarray}
  Each of these solutions can be extended to a full solution
  $\Gamma^{\rm exact}$ that contains the same physics and satisfies 
\begin{eqnarray}
\Sym(\Gamma^{\rm exact}) & = & 0\ ,\\
\Gamma|_{a=\chi=0} & = & \Gamma^{\rm exact}|_{a=\chi=0}\ .
\label{Extension}
\end{eqnarray}
  Since in the evaluation of $\Sym(\Gamma)|_{a=\chi=0}$ the unphysical
  parameters do not appear one has no need to calculate Feynman rules
  or vertex functions involving these parameters.
  This theorem is proven in subsec.\ \ref{Sec:ClassicalSolution} for
  the classical level and subsec.\ \ref{Sec:HigherOrders} for the
  quantum level.
\end{enumerate}
\def\thph{1{}}
\def\thsym{2{}}
For practical calculations the theorems have an important
implication. It is a possible and sufficient prescription to impose
only $\Sym(\Gamma)|_{a=\chi=0}=0$ and require normalization conditions
only for the physical parameters listed in theorem 1. Each solution of
this prescription is equivalent in physics respects to a full solution
of the symmetry identities, and any two solutions differ only in the
physically irrelevant part.

The proofs of these theorems are now given in the order of their
logical interdependence. First we prove a lemma which is a more
general form of theorem \thsym\ on the classical level and introduce
some useful notation. Then
this lemma is used to prove theorem \thph\ and finally theorem \thsym\
on the quantum level.

\subsection{Classical solution and invariant counterterms}
\label{Sec:ClassicalSolution}

Let $R$ be the following operator for a renormalization transformation
of all parameters and fields appearing in $\Gamma_{\rm cl,\
  canonical}|_{a=\chi=0}$ defined in eq.\ (\ref{GammaClGen}):
\begin{eqnarray}
\label{RenTransformation}
\lefteqn{{R}:}\nonumber\\
\{A^\mu, Y_{A^\mu}, & \to &
\{\sqrt{Z_A} A^\mu, \sqrt{Z_A}^{-1}Y_{A^\mu},\quad
\nonumber\\
B, \bar{c}, \xi\} &  & \quad\sqrt{Z_A}^{-1}B,
  \sqrt{Z_A}^{-1}\bar{c}, Z_A\xi\}
\nonumber\\
\{\lambda, Y_{\lambda}\} & \to &
\{\sqrt{Z_\lambda}\lambda, \sqrt{Z_\lambda}^{-1}\}
\nonumber\\
\{c, Y_c\} & \to & 
\{\sqrt{Z_c} c, \sqrt{Z_c}^{-1} Y_c\}
\nonumber\\
\{\phi_i, Y_{\phi_i}\} & \to & 
\{\sqrt{Z_\phi}_{ij}\phi_j, \sqrt{Z_\phi}^{-1}_{ij}Y_{\phi_j}\}
\nonumber\\
\{\psi_i, Y_{\psi_i}\} & \to & 
\{\sqrt{Z_\psi}_{ij}\psi_j, \sqrt{Z_\psi}^{-1}_{ij}Y_{\psi_j}\}
\nonumber\\
\{g, m_{ij}, g_{ijk}\} & \to &
\{g+\delta g, m_{ij}+\delta m_{ij}, g_{ijk}+\delta g_{ijk}\}
\nonumber\\
\{\tilde{M}^2_{ij}, \tilde{B}_{ij}, \quad & \to &
\{\tilde{M}^2_{ij}+\delta \tilde{M}^2_{ij}, 
\tilde{B}_{ij}+\delta \tilde{B}_{ij}, 
\nonumber\\
\tilde{A}_{ijk},
\tilde{M}_\lambda\} & &
\quad\tilde{A}_{ijk}+\delta \tilde{A}_{ijk}, 
\tilde{M}_\lambda+\delta \tilde{M}_\lambda\}
\nonumber\\
\{u_{3ij}(0), v_{3}(0)\} & \to & 
\{u_{3ij}(0)+\delta u_{3ij}(0), v_{3}(0)+\delta v_{3}(0)\}
\nonumber\\
\label{DefR}
\end{eqnarray}
with real constants $\sqrt{Z_A}$, $\sqrt{Z_\lambda}$, 
$\sqrt{Z_c}$, $\sqrt{Z_\phi}{}_{ij}$, $\sqrt{Z_\psi}{}_{ij}$, 
$\delta g$, $ \delta m_{ij}$, $ \delta g_{ijk}$, $\delta \tilde{M}^2_{ij}$, 
$ \delta \tilde{B}_{ij}$, $ \delta\tilde{A}_{ijk}$, 
$ \delta \tilde{M}_\lambda$, $ \delta u_{3ij}(0)$, 
$ \delta v_{3}(0)$ that have to be compatible with the global
symmetries. 

Similarly, let $\delta R$ be the following infinitesimal
renormalization transformation: 
\begin{eqnarray}
\lefteqn{\delta R =
  \frac12\delta Z_A \Bigl[\intx\Bigl(
    \Number{A_a^\mu} - \Number{Y_{A^\mu_a}}
}
 \nonumber\\&&{}\qquad
  - \Number{B_a} - \Number{\bar{c}_a}\Bigr) 
  + 2\xi\dpartial{}{\xi}\Bigr]
 \nonumber\\&&{}
 + \frac12\delta Z_\lambda \intx \Bigl(
    \Number{\lambda_a} + \Number{\lambdabar_a}
 \nonumber\\&&{}\qquad
  - \Number{Y_{\lambda_a}} - \Number{Y_{\lambdabar_a}}\Bigr)
 \nonumber\\&&{}
 + \frac12\delta Z_c \intx\Bigl(
    \Number{c} - \Number{Y_c}\Bigr)
 \nonumber\\&&{}
 + \frac12\delta Z_{\phi}{}_{ij}\intx  \Bigl(
    \phi_j\dfunc{}{\phi_i} + \phi_j^\dagger\dfunc{}{\phi_i^\dagger}
 \nonumber\\&&{}\qquad
  - Y_{\phi_i}\dfunc{}{Y_{\phi_j}}
  - Y_{\phi_i^\dagger}\dfunc{}{Y_{\phi_j^\dagger}}
 \Bigr)
 \nonumber\\&&{}
 + \frac12\delta Z_{\psi}{}_{ij}\intx \Bigl(
    \psi_j^\alpha\dfunc{}{\psi_i^\alpha}
     + \psibar_j{}_\alphadot\dfunc{}{\psibar_i{}_\alphadot}
 \nonumber\\&&{}\qquad
  - Y_{\psi_i}^\alpha\dfunc{}{Y_{\psi_j}^\alpha}
  - Y_{\psibar_i}{}_\alphadot\dfunc{}{Y_{\psibar_j}{}_\alphadot}
 \Bigr)
 \nonumber\\&&{}
 + \delta g \dpartial{}{g}
 + \delta m_{ij} \dpartial{}{m_{ij}}
 + \delta g_{ijk} \dpartial{}{g_{ijk}}
 \nonumber\\&&{}
 + \delta \tilde{M}_{ij}^2 \dpartial{}{\tilde{M}_{ij}^2}
 + \delta \tilde{B}_{ij} \dpartial{}{\tilde{B}_{ij}}
 + \delta \tilde{A}_{ijk} \dpartial{}{\tilde{A}_{ijk}}
 + \delta \tilde{M}_{\lambda} \dpartial{}{\tilde{M}_{\lambda}}
 \nonumber\\&&{}
 + \delta u_{3ij}(0) \dpartial{}{u_{3ij}(0)}
 + \delta v_3(0)\dpartial{}{v_3(0)} 
 \ .
\label{DefDeltaR}
\end{eqnarray}
According to the results of sec.~\ref{Sec:GenClassicalSolution} and using
the identification
\begin{eqnarray}
\sqrt{Z_\phi}{}_{ij} & \to & u_{1ij}\ ,\nonumber\\
\sqrt{Z_\psi}{}_{ij} & \to & (u_1u_2)_{ij}\ ,\nonumber\\
\sqrt{Z_A} & \to & v_{1}\ ,\nonumber\\
\sqrt{Z_\lambda} & \to & v_{1}v_2\ ,
\end{eqnarray}
we see that both operators $R,\delta R$ are compatible with the
symmetries. Suppose, $\Gamma_{\rm cl}$ is a classical solution of
$\Sym(\Gamma_{\rm cl})=0$. Then $R\Gamma_{\rm cl}$ is another
solution: 
\begin{eqnarray}
\Sym(R\Gamma_{\rm cl}) & = & 0\ ,
\label{RGammaCl}
\end{eqnarray}
and $\delta R$ generates symmetric counterterms (compare eq.\
(\ref{ClSym})):
\begin{eqnarray}
\Gamma_{\rm sym} & = & \delta R\Gamma_{\rm cl}\nonumber\\
\Rightarrow\quad
\Sym(\Gamma_{\rm cl}+\zeta\Gamma_{\rm sym}) & = & 0 +
{\cal O}(\zeta^2)\ .
\label{RGammaSym}
\end{eqnarray}

Now we consider the symmetry identities and its classical solutions in
the limit 
\begin{eqnarray}
&&a=\chi=0, f\mbox{ arbitrary.}
\label{IntermediateLimit}
\end{eqnarray}
This limit is not identical with the physical limit
(\ref{PhysicalLimit}) but better suited for our needs. In this limit
the unwanted parameters do not 
appear but still the symmetry identities are restrictive enough.
\paragraph{Lemma:}
Let $\Gamma_{\rm cl}$ and $\Gamma_{\rm sym}$ denote a classical
solution and an action for symmetric counterterms in the limit
$a=\chi=0$,
\begin{eqnarray}
\Sym(\Gamma_{\rm cl})|_{a=\chi=0} & = & 0\ ,
\label{IdSym1}
\\
\Sym(\Gamma_{\rm cl}+\zeta\Gamma_{\rm sym})|_{a=\chi=0} & = & 0 +
{\cal O}(\zeta^2)\ .
\label{IdSym2}
\end{eqnarray}
Then the most general form of $\Gamma_{\rm cl}$,
$\Gamma_{\rm sym}$ has to fulfil the relations
\begin{eqnarray}
\label{Lemma1}
\Gamma_{\rm cl}|_{a=\chi=0} & = & 
[R\Gamma_{\rm cl,\ canonical}]|_{a=\chi=0}\ ,\\
\Gamma_{\rm sym}|_{a=\chi=0} & = & 
[\delta R\Gamma_{\rm cl,\ canonical}]|_{a=\chi=0}\ ,
\label{Lemma2}
\end{eqnarray}
with the operators $R$, $\delta R$ defined in (\ref{DefR}),
(\ref{DefDeltaR}). 
\paragraph{Proof:}
The general classical solution of the symmetry identities
(\ref{IdSym1}), (\ref{IdSym2}) can be obtained by a straightforward
calculation. We write down a general ansatz, apply the symmetry
identities and derive the necessary relations the coefficients in the
ansatz have to satisfy. Although the calculation is lengthy, the
announced results (\ref{Lemma1}), (\ref{Lemma2}) follow in a direct
way.  

We now give a short sketch of the calculation with emphasis on the
main point, namely the restriction of the terms of ${\cal
  O}(\hat{f},\hat{f}^\dagger)$. This sketch will also show why we have
to use the limit (\ref{IntermediateLimit}) instead of
(\ref{PhysicalLimit}) in the statement of the lemma. 

The most general ansatz for
$\Gamma_{\rm cl}$ can be decomposed according to the degree in
$a,\chi,\hat{f}$: 
\begin{eqnarray}
\Gamma_{\rm cl} & = & \Gamma_{0} + \Gamma_{\hat{f},\rm\,lin}
 + \Gamma_{\hat{f},\rm\,rest} + \Gamma_{\chi,\rm\,lin}
 + \Gamma_{\rm rest}\ ,
\end{eqnarray}
where $\Gamma_0$ does not depend on $a,\chi,\hat{f}$;
$\Gamma_{\hat{f},\rm\,lin}, \Gamma_{\hat{f},\rm\,rest}$ are linear and
of higher degree in $\hat{f}$ but do not depend on
$a,\chi$; $\Gamma_{\chi,\rm\,lin}$ is linear in $\chi$ and does not
depend on $a$, $\hat{f}$, and
$\Gamma_{\rm rest}$ contains the rest of the dependence on $\chi$,
$\hat{f}$, and the complete dependence on $a$.

Since all defining symmetry identities either do not change the degree
in $a,\chi,\hat{f}$ or increase it, we obtain for $\Gamma_0$:
\begin{eqnarray}
0 & = & \Sym(\Gamma)|_{a=\chi=\hat{f}=0} = \Sym(\Gamma_0)\ ,
\end{eqnarray}
thus $\Gamma_0$ is a classical solution of the defining symmetry
identities in the case without soft breaking \cite{MPW96a}.

Next, the symmetry identities in (\ref{IdSym1}) imply that
  $\Gamma_{\hat{f},\rm\ lin}$ is globally invariant and does not
  depend on $B_a$ and $\omega^\mu$, and that
\begin{eqnarray}
0 & = & S(\Gamma)|_{a=\chi=0,\mbox{\scriptsize\,linear in }\hat{f}}
\nonumber\\
  & = & s_{\Gamma_0}^0\
\Gamma_{\hat{f},\rm\,lin} + S_\chi(\Gamma_{\chi,\rm\,lin})\ .
\end{eqnarray}
Here $s^0_{\Gamma_0}$ is the linearized version of $S_0$ defined by
\begin{eqnarray}
S_0(\Gamma_0+\zeta\Gamma_1) & = & S_0(\Gamma_0) + \zeta s^0_{\Gamma_0}
\Gamma_1 + {\cal O}(\zeta^2)\ ,
\end{eqnarray}
and
\begin{eqnarray}
S_\chi(\Gamma) & = & \intx\Bigl(
s\chi^\alpha\dfrac{\Gamma}{\chi^\alpha}\Big|_{a=\chi=0}
+s\chibar_\alphadot\dfrac{\Gamma}{\chibar_\alphadot}\Big|_{a=\chi=0}
\Bigr)
\nonumber\\& = & \intx\Bigl(
\sqrt2\hat{f}\epsilon^\alpha\dfrac{\Gamma}{\chi^\alpha}\Big|_{a=\chi=0}
\nonumber\\&&{}\qquad
-\sqrt2\hat{f}^\dagger\epsilonbar_\alphadot
\dfrac{\Gamma}{\chibar_\alphadot}\Big|_{a=\chi=0}\Bigr)\ .
\label{DefSChi}
\end{eqnarray}
Due to the form of the operator $S_\chi$ we obtain
\begin{eqnarray}
s_{\Gamma_0}^0\ \Gamma_{\hat{f},\rm\,lin} &=&
{\cal O}(\epsilon\hat{f})+{\cal O}(\epsilonbar\hat{f}^\dagger)\ .
\label{LemmaEq1}
\end{eqnarray}
Since on the physical fields $s^0_{\Gamma_0}$ acts as the BRS operator
$s$ up to field and parameter renormalizations, it is easy to see that
the most general solution for $\Gamma_{\hat{f},\rm\,lin}$ that is
compatible with the global symmetries is given
by
\begin{eqnarray}
\Gamma_{\hat{f},\rm\,lin} & = &
\hat{f}\Bigl(\tilde{A}_{ijk}\phi_i\phi_j\phi_k +
\tilde{B}_{ij}\phi_i\phi_j
 + \tilde{M}_\lambda \lambda_a\lambda_a
\nonumber\\&&{}
 + u_{3ij}\sqrt2 Y_{\psi_i}\epsilon \hat{f}\phi_j + v_3\sqrt2
 Y_{\lambdabar_a}\sigmabar^\mu\epsilon A_{a\mu}\Bigr)
\nonumber\\&&{}+h.c.
\end{eqnarray}
All these terms are accounted for in the operator $R$, eq.\
(\ref{RenTransformation}). 

This is the point where the limit (\ref{IntermediateLimit}) is
important. If we had required only $\Sym(\Gamma_{\rm
  cl})|_{a=\chi=f=0}$ instead of eq.\ (\ref{IdSym1}), then we would
have obtained only ${\cal O}(\epsilon)+{\cal O}(\epsilonbar)$ on the
r.h.s.\ of eq.\ (\ref{LemmaEq1}), and in the solution to this
equation non-GG terms $\phi\phi\phi^\dagger$ or $\psi\psi$ would have
appeared. 

The constraints on the remaining parts of $\Gamma_{\rm cl}$ can be
worked out similarly.

\subsection{Physical parameters}

\label{Sec:PhysParameters}

Once the symmetry identities are satisfied at a given order in the
limit (\ref{IntermediateLimit}), there can still be divergent
contributions which have to be absorbed by symmetric counterterms
$\Gamma_{\rm sym}$ 
satisfying 
\begin{eqnarray}
\Sym(\Gamma_{\rm cl} + \zeta\Gamma_{\rm sym})|_{a=\chi=0} & = &
0 + {\cal O}(\zeta^2)\ .
\label{GammaSym}
\end{eqnarray}
According to the lemma the most general form of $\Gamma_{\rm sym}$
is generated by the infinitesimal renormalization transformation 
\begin{eqnarray}
\Gamma_{\rm sym}|_{a=\chi=0} & = & [\delta {R}\Gamma_{\rm
  cl}]|_{a=\chi=0} \ .
\end{eqnarray}

This leads to the following hierarchy of the symmetric counterterms:
\begin{enumerate}
\item Counterterms appearing in physical processes, where not only
  $a=\chi=0$, but also the external $Y_i$ fields are set to zero:
\begin{eqnarray}
\Gamma_{\rm sym}|_{a=\chi=0,    Y_i=0}\ .
\end{eqnarray}
This first class contains the counterterms to the field
renormalization constants $Z_A, Z_\lambda, Z_c, Z_\phi, Z_\psi$ and
the parameters $g, m_{ij}, g_{ijk}$, $\tilde{M}^2_{ij},
\tilde{B}_{ij}, \tilde{A}_{ijk}, \tilde{M}_\lambda$. 
\item Additional counterterms appearing for $Y_i\ne0$:
\begin{eqnarray}
\Gamma_{\rm sym}|_{a=\chi=0,    Y_i\ne0}\ .
\end{eqnarray}
This class contains precisely the counterterms to the $u_3, v_3$
parameters. 
\item The rest of the counterterms appearing for $a,\chi$ arbitrary:
\begin{eqnarray}
\Gamma_{\rm sym}|_{a,\chi\ne0,    Y_i\ne0}\ .
\end{eqnarray}
This class contains infinitely many independent counterterms.
\end{enumerate}

The normalization conditions fixing the first, second and third set of
counterterms we call {\em normalization conditions of the first,
second and third class}, respectively.

The next theorem states how far we get using only the
class-one-normalization conditions and leaving open the ones of the
second and third class.
\paragraph{Theorem \thph:}
Two solutions $\Gamma_1$ and $\Gamma_2$ of the same class-one-normalization
conditions and of the symmetry identities in the limit 
(\ref{IntermediateLimit}),
\begin{eqnarray}
\Sym(\Gamma_2)|_{a=\chi=0} = \Sym(\Gamma_1)|_{a=\chi=0} & = & 0\ ,
\end{eqnarray}
can differ at most by local terms proportional to $Y_\psi, Y_\lambda$:
\begin{eqnarray}
\lefteqn{(\Gamma_2-\Gamma_1)|_{a=\chi=0}}
\nonumber\\
 & = &
\Delta_Y(u_{3ij}(0)+\delta u_{3ij}(0),v_3(0)+\delta v_3(0))
\nonumber\\
& \equiv & \intx\Bigl(
-Y_{\psi_i}{}_\alpha
       \sqrt2\epsilon^\alpha \hat{f}(u_{3ij}(0)+\delta u_{3ij}(0)) \phi_j
\nonumber\\&&{}
- Y_{\lambda}{}_\alpha \sqrt2(v_3(0)+\delta v_3(0))\hat{f}^\dagger
\epsilonbar\sigmabar^\mu{}^\alpha A_\mu\Bigr)
+h.c.
\end{eqnarray}

\paragraph{Proof:}
Due to the lemma this holds at the tree level. To perform an
inductive proof of this statement we suppose that we have at the order
$\hbar^{n-1}$:
\begin{eqnarray}
\label{ClaimStart}
(\Gamma_2-\Gamma_1)|_{a=\chi=0} & = & \Delta_Y(u_3^{(n-1)},v_3^{(n-1)})
\nonumber\\&&{}
+{\cal O}(\hbar^n)
\ ,\\
(\Gamma_{2,\rm ct}-\Gamma_{1,\rm ct})|_{a=\chi=0}
 & = & \Delta_Y(\delta u_3^{(n-1)},\delta v_3^{(n-1)})
\nonumber\\&&{}
+{\cal O}(\hbar^n)
\ .
\label{ClaimEnd}
\end{eqnarray}
Then, at the next order {\em all} one-particle irreducible loop
diagrams not involving $a,\chi$ are the same, regardless whether
calculated according to the Feynman rules for $\Gamma_1$ or
$\Gamma_2$. This is true because even though the Feynman rules differ by
the terms $\Delta_Y$, these differences cannot contribute since they
are linear in the propagating fields. 

The difficult point is to prove that the counterterms of the order
$\hbar^n$, denoted by $\Gamma_{\rm 1,\ ct}^{(n)}$ and 
$\Gamma_{\rm 2,\ ct}^{(n)}$, do not invalidate
(\ref{ClaimStart}-\ref{ClaimEnd}). We know
\begin{eqnarray}
(\Gamma_2-\Gamma_1)|_{a=\chi=0} & = & \Delta\Gamma^{(n)}_{\rm ct} + 
\Delta_Y(u_3^{(n-1)},v_3^{(n-1)})
\nonumber\\&&{}
 + {\cal O}(\hbar^{n+1})    \ ,
\label{DiffNextOrder}
\\
\Delta\Gamma^{(n)}_{\rm ct} & = & 
(\Gamma^{(n)}_{2,\ \rm ct}-\Gamma^{(n)}_{1,\ \rm ct})|_{a=\chi=0}\ .
\end{eqnarray}
Thus, taking into account the symmetry of $\Delta_Y$ and the fact that
all symmetry identities except for the Slavnov-Taylor
identity are linear and do not change the degree in $a,\chi$, we
obtain for these identities
\begin{eqnarray}
0 & = & \Sym(\Gamma_2)|_{a=\chi=0}\nonumber\\
  & = & \Sym(\Gamma_2|_{a=\chi=0})\nonumber\\
  & = & \Sym(\Gamma_1|_{a=\chi=0}+\Delta\Gamma^{(n)}_{\rm ct}
\nonumber\\&&{}
        + \Delta_Y(u_3^{(n-1)},v_3^{(n-1)}))\nonumber\\
  & = & 0 + \Sym(\Delta\Gamma^{(n)}_{\rm ct})\ .
\end{eqnarray}
For the Slavnov-Taylor identity we obtain at the order $\hbar^n$ (we
use the operator $S_\chi$ defined in eq.\ (\ref{DefSChi})):
\begin{eqnarray}
0& = & S(\Gamma_{2})|_{a=\chi=0}
\nonumber\\
 & = & S_0(\Gamma_{2}|_{a=\chi=0})
     + S_\chi(\Gamma_{2})
\nonumber\\
 & = & S_0(\Gamma_1|_{a=\chi=0}+\Delta\Gamma^{(n)}_{\rm ct}+\Delta_Y)
     + S_\chi(\Gamma_{2})
\nonumber\\
 & = & S(\Gamma_1+\Delta\Gamma^{(n)}_{\rm ct}+\Delta_Y)|_{a=\chi=0}
\nonumber\\&&{}
     + S_\chi(\Gamma_{2}-(\Gamma_1+\Delta\Gamma^{(n)}_{\rm ct}+\Delta_Y))
\nonumber\\
 & = & S(\Gamma_1+\Delta\Gamma^{(n)}_{\rm ct})|_{a=\chi=0} 
\nonumber\\&&{} +
 \intx\Bigl(\dfrac{\Gamma_1+\Delta\Gamma^{(n)}_{\rm ct}}{Y_i}
            \dfrac{\Delta_Y}{\varphi_i} 
\nonumber\\&&{}\qquad
           +\dfrac{\Delta_Y}{Y_i}
            \dfrac{\Gamma_1+\Delta\Gamma^{(n)}_{\rm ct}}{\varphi_i}
\Bigr)|_{a=\chi=0}
\nonumber\\&&{}
     + S_\chi(\Gamma_{2}-(\Gamma_1+\Delta\Gamma^{(n)}_{\rm ct}+\Delta_Y))
\nonumber\\
 & = & S(\Gamma_1+\Delta\Gamma^{(n)}_{\rm ct})|_{a=\chi=0} 
\nonumber\\&&{}
+\sqrt2(\epsilon^\alpha X_\alpha\hat{f}
 -\epsilonbar_\alphadot\bar{X}^\alphadot\hat{f}^\dagger)\ .
\label{Last}
\end{eqnarray}
The last two equations hold owing to the special form of $\Delta_Y$
with some suitably chosen functional $X_\alpha$. 
Since $\Gamma_1$ satisfies the Slavnov-Taylor identity the first term
of this result can be simplified using
\begin{eqnarray}
S(\Gamma_1+\Delta\Gamma^{(n)}_{\rm ct}) & = &
S(\Gamma_{1,\rm cl}+\Delta\Gamma^{(n)}_{\rm ct})+{\cal O}(\hbar^{n+1})
\ .
\end{eqnarray}
Therefore, both terms in the last line of eq.\ (\ref{Last}) are local
and power-counting renormalizable functionals of the order $\hbar^n$,
and we can define a counterterm action
\begin{eqnarray}
\Gamma_{\rm sym} & = & \Delta\Gamma^{(n)}_{\rm ct} 
                        + (\chi^\alpha X_\alpha +
                        \chibar_\alphadot\bar{X}^\alphadot) 
\end{eqnarray}
that satisfies
\begin{eqnarray}
S(\Gamma_{1,\rm cl}+\Gamma_{\rm sym})|_{a=\chi=0} & = & 
S(\Gamma_1+\Delta\Gamma^{(n)}_{\rm ct})|_{a=\chi=0} 
\nonumber\\&&{}
+\sqrt2(\epsilon^\alpha X_\alpha\hat{f}
 -\epsilonbar_\alphadot \bar{X}^\alphadot\hat{f}^\dagger)
\nonumber\\
& = & 0 + {\cal O}(\hbar^{n+1})
\ .
\end{eqnarray}
Thus, $\Gamma_{\rm sym}$ is a symmetric counterterm in the sense of
eq.~(\ref{GammaSym}), and we obtain from the lemma:
\begin{eqnarray}
\Gamma_{\rm sym}|_{a=\chi=0} & = & [\delta R\Gamma_{1,\rm
  cl}]|_{a=\chi=0} 
\end{eqnarray}
On the other hand, by construction $\Gamma_{\rm sym}$ contains the
relevant difference of $\Gamma_1$ and $\Gamma_2$ at the order
$\hbar^n$: 
\begin{eqnarray}
(\Gamma_2-\Gamma_1)|_{a=\chi=0} & = & \Gamma_{\rm sym}|_{a=\chi=0} + 
\Delta_Y(u_3^{(n-1)},v_3^{(n-1)})
\nonumber\\&&{}
 + {\cal O}(\hbar^{n+1})\ .
\end{eqnarray}
Now, since $\Gamma_{1,2}$ satisfy the same class-one-nor\-ma\-li\-za\-tion
conditions, $\Gamma_{\rm sym}$ cannot contain any
class-one-coun\-ter\-terms. Since these are the only counterterms that
appear in the limit $a=\chi=Y_i=0$, we obtain
\begin{eqnarray}
\Gamma_{\rm sym}|_{a=\chi=Y_i=0} & = & 0\ .
\end{eqnarray}
Owing to the concrete form of $\delta R$, this shows
\begin{eqnarray}
\Delta\Gamma_{\rm ct}^{(n)}|_{a=\chi=0} & = & 
\Gamma_{\rm sym}|_{a=\chi=0} = 
\Delta_Y(\delta u_3^{(n)},\delta v_3^{(n)})
\ .
\end{eqnarray}
Together with eq.\ (\ref{DiffNextOrder}) this demonstrates the
validity of eqs.\ (\ref{ClaimStart}-\ref{ClaimEnd}) at the next order,
completing the induction. 

\subsection{Simplified symmetry identities at the quantum level}
\label{Sec:HigherOrders}

According to theorem \thph, the parameters of class 2 and class 3
are irrelevant in physics respects. In this subsection, a
complementary theorem is proven. This theorem \thsym\ states that it
is sufficient to establish the symmetry identities in the limit
(\ref{IntermediateLimit}), where the infinitely many parameters of
class 3 do not appear at all. This implies that the class 3 parameters
can be completely ignored in practice. The two parameters $u_3$, $v_3$
of class 2 are also unphysical, but they do appear in the limit
(\ref{IntermediateLimit}).

At the classical level, this is a direct consequence of the Lemma in
subsec.\ \ref{Sec:ClassicalSolution} together with 
eqs.\ (\ref{RGammaCl}), (\ref{RGammaSym}): Any classical solution
$\Gamma_{\rm cl}$ of the symmetry identities (\ref{IdSym1}) is
equivalent in physics respects to a solution $[R\Gamma_{\rm cl,\
  canonical}]$ of the full symmetry identities. In this subsection 
the theorem is extended to the quantum level. The statement of the
theorem and its proof at the quantum level is divided into two
parts---the existence of a solution to the symmetry identities in the
limit (\ref{IntermediateLimit}) and its extension to a full solution.

\subsubsection{Existence of a solution}

\def\thqua{\thsym{}a{}}
\paragraph{Theorem \thqua:}
Suppose $\Gamma$ is a solution of the symmetry identities in the limit
(\ref{IntermediateLimit}) up to the order $\hbar^{n-1}$,
\begin{eqnarray}
\Sym(\Gamma)|_{a=\chi=0} & = & 0 + {\cal O}(\hbar^n)\ ,
\label{ClaimSym}
\end{eqnarray}
and $\Gamma^{\rm exact}$ is an extension that solves the full symmetry
identities,
\begin{eqnarray}
\Sym(\Gamma^{\rm exact}) & = & 0 + {\cal O}(\hbar^n)\ ,\\
(\Gamma^{\rm exact}-\Gamma)|_{a=\chi=0} & = & 0 + {\cal O}(\hbar^n)\ .
\label{ClaimEq}
\end{eqnarray}
Then we claim that $\Gamma$, $\Gamma^{\rm exact}$ can be renormalized
in such a way that the eqs.\ (\ref{ClaimSym}-\ref{ClaimEq}) are
maintained at the next order $\hbar^n$.

\paragraph{Proof:} 
Since we assume the absence of anomalies, $\Gamma^{\rm exact}$ can be
renormalized in such a way that 
\begin{eqnarray}
\Sym(\Gamma^{\rm exact}) & = & 0 + {\cal O}(\hbar^{n+1})\ .
\end{eqnarray}
Since the Feynman rules of the order $\hbar^n$ defined by $\Gamma^{\rm
  exact}$ and $\Gamma$ differ only in terms $\sim a,\chi$, all loop
diagrams contributing to $\Gamma^{\rm exact}|_{a=\chi=0}$ and 
$\Gamma|_{a=\chi=0}$ are equal at this order. Thus, adding appropriate
${\cal O}(\hbar^n)$ counterterms to $\Gamma$ we obtain
\begin{eqnarray}
(\Gamma^{\rm exact}-\Gamma)|_{a=\chi=0} & = & 0 + {\cal
  O}(\hbar^{n+1})\ .
\label{ProofEq}
\end{eqnarray}
However, $\Gamma$ does not yet satisfy the Slavnov-Taylor identity at
this order. Indeed, neglecting terms of the order $\hbar^{n+1}$ we
obtain 
\begin{eqnarray}
S(\Gamma)|_{a=\chi=0} & = & S_0(\Gamma|_{a=\chi=0})
                           +S_\chi(\Gamma)
\nonumber\\
& = & S_0(\Gamma^{\rm exact}|_{a=\chi=0})
     +S_\chi(\Gamma)
\nonumber\\& = & S(\Gamma^{\rm exact})|_{a=\chi=0}
                +S_\chi(\Gamma-\Gamma^{\rm exact})
\nonumber\\
& = & S_\chi(\Gamma-\Gamma^{\rm exact})
\nonumber\\
& = & \hbar^n\Delta
\ .
\end{eqnarray}
Owing to  the quantum action principle
\cite{QAP}, the lowest order of $\Delta$ is a local and power-counting
renormalizable functional, and owing to the form of $S_\chi$ it takes
the form
\begin{eqnarray}
\Delta & = & \int \sqrt2\epsilon^\alpha X_\alpha \hat{f} 
          - \sqrt2\epsilonbar_\alphadot \bar{X}^\alphadot \hat{f}^\dagger 
+ {\cal O}(\hbar)
\ .
\end{eqnarray}
Hence, adding the counterterms 
\begin{eqnarray}
\Gamma\to\Gamma-\int \hbar^n(\chi^\alpha X_\alpha
                     +\chibar_\alphadot\bar{X}^\alphadot)
\end{eqnarray}
restores the Slavnov-Taylor identity $S(\Gamma)|_{a=\chi=0}=0+{\cal
  O}(\hbar^{n+1})$ without interfering with eq.\
(\ref{ProofEq}). All further symmetry identities are linear and
homogeneous in $a,\chi$. Therefore, $\Gamma$ satisfies these
identities, too, and we obtain
\begin{eqnarray}
\Sym(\Gamma)|_{a=\chi=0} & = & 0 + {\cal O}(\hbar^{n+1})\ .
\end{eqnarray}
This was to be shown.

\subsubsection{Extension to a full solution}

\def\thqub{\thsym{}b{}}
\paragraph{Theorem \thqub:}
Let $\Gamma$ be a solution to the symmetry identities in the limit
$a=\chi=0$,
\begin{eqnarray}
\label{SymGammaTh2}
\Sym(\Gamma)|_{a=\chi=0} & = & 0\ .
\end{eqnarray}
Then there exists an extension to a full solution $\Gamma^{\rm exact}$
satisfying 
\begin{eqnarray}
\label{Claim2bStart}
\Sym(\Gamma^{\rm exact}) & = & 0\ ,\\
(\Gamma^{\rm exact}-\Gamma)|_{a=\chi=0} & = & 0\ .
\label{Claim2bEnd}
\end{eqnarray}

\paragraph{Proof:}
Due to the lemma there is a classical solution $\Gamma^{\rm
exact}_{\rm cl}$ satisfying eqs.\
(\ref{Claim2bStart}-\ref{Claim2bEnd}). Now suppose the same is true at
the order $\hbar^{n-1}$, that is there exists an effective action
$\Gamma^{\rm exact}$ satisfying
\begin{eqnarray}
\Sym(\Gamma^{\rm exact}) & = & 0+{\cal O}(\hbar^n)\ ,\\
(\Gamma^{\rm exact}-\Gamma)|_{a=\chi=0} & = & 0+{\cal O}(\hbar^n)\ .
\end{eqnarray}
Then, according to theorem \thqua\ there are ${\cal O}(\hbar^n)$
counterterms yielding $\widetilde{\Gamma}=\Gamma+{\cal O}(\hbar^n)$,
$\widetilde{\Gamma}^{\rm exact}=\Gamma^{\rm exact}+{\cal O}(\hbar^n)$
such that
\begin{eqnarray}
\label{SymGammaTilde}
\Sym(\widetilde{\Gamma})|_{a=\chi=0} & = & 0+{\cal O}(\hbar^{n+1})\ ,\\
\Sym(\widetilde{\Gamma}^{\rm exact}) & = & 0+{\cal O}(\hbar^{n+1})\ ,\\
(\widetilde{\Gamma}^{\rm exact}-\widetilde{\Gamma})|_{a=\chi=0} 
& = & 0+{\cal O}(\hbar^{n+1})\ .
\end{eqnarray}
However, due to eqs.\ (\ref{SymGammaTh2}), (\ref{SymGammaTilde}) the
difference $\widetilde{\Gamma}-\Gamma$ has to be a symmetric
counterterm as defined in eq.\ (\ref{IdSym2}). Hence, it has the form
\begin{eqnarray}
(\Gamma-\widetilde{\Gamma})|_{a=\chi=0} & = & [\delta R\Gamma_{\rm cl}]
|_{a=\chi=0}\ .
\end{eqnarray}
Therefore, $\Gamma^{\rm exact}=\widetilde{\Gamma}^{\rm exact}+\delta
R\Gamma^{\rm exact}_{\rm cl}$ has the desired properties
\begin{eqnarray}
\Sym(\Gamma^{\rm exact}) & = & 
\Sym(\widetilde{\Gamma}^{\rm exact}+\delta
R\Gamma^{\rm exact}_{\rm cl})\nonumber\\
& = & 0+{\cal O}(\hbar^{n+1})\ ,\\
(\Gamma^{\rm exact}-\Gamma)|_{a=\chi=0} & = & 
(\widetilde{\Gamma}^{\rm exact}-\widetilde{\Gamma})|_{a=\chi=0}
\nonumber\\
& = & 0+{\cal O}(\hbar^{n+1})\ .
\end{eqnarray}
This completes the induction.

\section{Alternative approach}
\label{Sec:Alternatives}

The first Slavnov-Taylor
identity for softly broken
supersymmetric gauge theories was presented in ref.~\cite{MPW96b}. In
this construction the absence of anomalies could be nicely shown, but
there appeared new kinds of parameters whose physical meaning remained
unclear. As shown in sec.\ \ref{Sec:PhysicalPart}, in our approach
this problem could be solved. In this section a brief comparison of
both approaches is given. 

Basically, in both approaches the soft
breaking is introduced via external fields with definite BRS
transformation rules. These transformation rules contain a constant
shift that yields the soft parameters in the limit of vanishing
external fields. 

The main difference concerns the underlying intuition and consequently
the external field content:\footnote{One further difference
  concerns the supersymmetric mass terms which 
  are also introduced via external fields in \cite{MPW96b}. This is
  done in order not to violate $R$-invariance because the $R$-weights
  of the chiral fields are fixed to $n_i=\frac23$ (translated to 
  our convention) in accordance with the $R$-part of the
  supercurrent. In our case the $R$-weights are assumed to be chosen
  in such a way that the mass terms are invariant and therefore we do
  not need such an external field multiplet.} 
In \cite{MPW96b}, the soft breaking terms are not introduced as
couplings to a multiplet 
$(a,\chi,\hat{f})$ that transforms as a chiral supermultiplet but as
couplings to a BRS doublet $(u,\hat{v})$ where\footnote{The equations
  are translated to
  our conventions. In particular, in \cite{MPW96b} there is also an
  $R$-transformation part in the BRS transformations, which is neglected
  here.}  
\begin{eqnarray}
s u & = & \hat{v} - i\omega^\nu\partial_\nu u\ ,\\
s v & = & 2i\epsilon\sigma^\nu\epsilonbar\partial_\nu u
                  -i\omega^\nu\partial_\nu v\ ,\\
\hat{v}(x) & = & v(x) + \kappa\ .
\end{eqnarray}
The main benefit of this structure is that the cohomological sector of
the theory is not altered compared to the case without soft
breaking. This allows a straightforward proof of the absence of
anomalies. 

Contrary to the case of $(a,\chi,\hat{f})$, however, the BRS
transformations of $u$ and $v$ cannot be interpreted as supersymmetry
transformations where simply the transformation parameter has been
promoted to a ghost. Moreover, $u$ and $v$ are two scalar fields and
therefore cannot form a supersymmetry multiplet. 
Correspondingly, the restriction of the breaking terms to the ones of
the GG-class is done by requiring $R$-invariance with specially chosen
$R$-weights. In \cite{MPW96b}, requiring supersymmetry alone would not
suffice to forbid non-GG terms (see sec.\ \ref{Sec:SoftBreaking}). On
the one hand, this opens a way to perform the renormalization of
theories with arbitrary supersymmetry breaking. But on the other hand
the emphasized role of $R$-invariance might obstruct a deeper
understanding of softly broken supersymmetry  and its influence on
typical consequences of supersymmetry like non-renormalization
properties.

In the limit of vanishing external fields, the classical action in
both approaches reduces to the same soft breaking action but for
non-vanishing external fields in both cases new parameters appear: in
our case the ones discussed in section
\ref{Sec:GenClassicalSolution}, in the case of \cite{MPW96b} for
instance the parameters $\rho_2,\rho_4$ that appear in 
the terms
\begin{eqnarray}
\Gamma_{2,4} & = & \intx\Bigl(\rho_{2ab}
Y_{\psi_b}{}_{\alpha}\epsilon^\alpha(\hat{v}\phi_a 
                     -\sqrt2 u\epsilon\psi_a)
\nonumber\\&&{}
+ \rho_4{}_{ab}\hat{v}\bar{u}\epsilon\psi_a\phi_b^\dagger
\Bigr)+\ldots
\label{Rho24}
\end{eqnarray}
The main reason why the approach of ref.\ \cite{MPW96b} cannot be used
directly in phenomenological applications is that the
physical meaning of these parameters is not obvious. In particular,
a theorem showing whether these parameters are irrelevant for physical
quantities or not---analogous to sec.\ \ref{Sec:PhysParameters}---is
lacking.

In spite of these differences, there is a remarkable relation between
both approaches. First of all, the quantum numbers of $\hat{v}$ and
$\hat{f}$ are equal, and second we can combine the supersymmetry
ghost and $u$ to a spinor $(\epsilon u)$ that has the same quantum
numbers as $\chi$. Hence, we can identify
\begin{eqnarray}
a&\to&0\ ,\nonumber\\
\chi^\alpha&\to&\epsilon^\alpha u\ ,\nonumber\\
\sqrt2\hat{f}&\to&\hat{v}\ .
\label{Identification}
\end{eqnarray}
Furthermore, this correspondence even holds for the BRS
transformations:
\begin{eqnarray}
sa&\to&\sqrt2\epsilon\epsilon u = 0\ ,\nonumber\\
s\chi^\alpha &\to& 
\sqrt2 \epsilon^\alpha\hat{v}-i\omega^\nu\partial_\nu
\epsilon^\alpha u = 
s\epsilon^\alpha u\ ,\nonumber\\ 
s\sqrt2\hat{f}&\to& 
2 i\epsilonbar\sigma^\nu\partial_\nu\epsilon u 
- i\omega^\nu\partial_\nu \hat{v} = s\hat{v}\ .
\label{BRSIdent}
\end{eqnarray}
Here we have used $\epsilon^\alpha\epsilon_\alpha=0$, which holds
since $\epsilon$ is bosonic. Thus, $u$ and $\hat{v}$ may be regarded
as a part of our chiral multiplet $(a,\chi,\hat{f})$. And there is a
natural identification in our framework of terms like the
$\rho_2$-term in (\ref{Rho24}), where $u$ comes always in combination
with $\epsilon$. In fact, this term has the same structure as the
$u_3$-term in eq.~(\ref{ModGammaExtPhiPsi}) with $u_3\to-\rho_2$ when 
(\ref{Identification}) is used. 

However, in the classical action of \cite{MPW96b} there are also terms
where $u$ appears without an accompanying $\epsilon$ or $\bar{u}$
without accompanying $\epsilonbar$, such as the $\rho_4$-term in
(\ref{Rho24}). These terms have no correspondence in 
our framework. On the other hand, of course our terms depending on the
$a$ field have no correspondence in \cite{MPW96b}. Therefore both
frameworks are really different and independent of each other.

%
%

\section{Conclusions}

In this article we have performed the renormalization of
supersymmetric Yang-Mills theories with soft su\-per\-sym\-me\-try-break\-ing
terms of the GG class. These terms are introduced in a supersymmetric
way via an external chiral multiplet, allowing a construction that
parallels the one without soft breaking.

This construction is afflicted by a problem, since in the course of
the renormalization,  an unconstrained number of additional
parameters appear. However, in sec.\ \ref{Sec:PhysicalPart} it is
shown that these parameters are irrelevant in physics respects. Even
better than gauge 
parameters they do not influence any vertex functions that occur in
physical S-matrix elements; and neither at the classical nor at the
quantum level it is necessary to calculate the part of the Lagrangian
and the counterterms involving those additional parameters. 

For practical calculations of physical processes the theorems in sec.\
\ref{Sec:PhysicalPart} imply, first,
that the symmetry identities need to be established only in the limit
(\ref{IntermediateLimit}), $\Sym(\Gamma)|_{a=\chi=0}=0$. And second,
renormalization of the fields and parameters appearing in the relevant
part of the classical action suffices to cancel the divergences.


Since the supersymmetric extensions of the standard model like the 
minimal one (MSSM) involve soft breaking, our results provide an
important building block for the renormalization of these kind of
models. 

The impossibility to accommodate non-GG breaking terms in the
framework with spurion fields, where breaking terms are introduced via
a coupling to a supermultiplet, shows that GG terms are a
renormalizable subclass of all breaking. That these terms have
even special properties under renormalization, as seen in explicit 
one-loop calculations and different approaches to their 
renormalization group coefficients \cite{Yamada,NRTGG,NRTs,SQEDNRT},
 cannot
be concluded 
by using the present formalism. As shown for the Abelian case
in \cite{SQEDNRT}, the present formalism provides
the correct starting point for this purpose, but it has to be 
enhanced by a deeper characterization of the symmetries.


\vspace{2ex}
{\em Acknowledgements.} We thank M.\ Roth, C.\ Rupp, and K.\ Sibold
for valuable discussions.

\begin{appendix}
\section{Conventions}

\paragraph{2-Spinor indices and scalar products:}

\begin{eqnarray}
\psi\chi & = & \psi^\alpha\chi_\alpha\ , 
\quad
\overline\psi\overline\chi  = 
\overline\psi_{\dot\alpha}\overline\chi^{\dot\alpha}\ ,
\end{eqnarray}

\paragraph{$\sigma$ matrices:}

\begin{eqnarray}\label{Paulimatrizen}
&&
\sigma^1  = \left(\begin{array}{cc}0&1\\1&0\end{array}\right),\quad
\sigma^2  = \left(\begin{array}{cc}0&-i\\i&0\end{array}\right),\quad
\sigma^3  = \left(\begin{array}{cc}1&0\\0&-1\end{array}\right) ,\quad\quad
\\
&&
\sigma^\mu_{\alpha\dot\alpha}  =  (1, \sigma^k)_{\alpha\dot\alpha}
\ ,\quad 
\overline\sigma^{\mu\dot\alpha\alpha}  =  (1,
-\sigma^k)^{\dot\alpha\alpha}
\ ,
\\&&
(\sigma^{\mu\nu})_\alpha\ ^\beta  = 
\frac{i}{2}(\sigma^\mu\overline\sigma^\nu-\sigma^\nu\overline\sigma^\mu)
_\alpha\ ^\beta
\ ,
\\&&
(\overline\sigma^{\mu\nu})^{\dot\alpha}\ _{\dot\beta}  = 
\frac{i}{2}(\overline\sigma^\mu\sigma^\nu-\overline\sigma^\nu\sigma^\mu)
^{\dot\alpha}\ _{\dot\beta}\ .
\end{eqnarray}

\paragraph{Complex conjugation:}

\begin{eqnarray}
(\psi\theta)^\dagger & = & \thetabar\psibar \ ,\\
(\psi\sigma^\mu\thetabar)^\dagger & = & \theta\sigma^\mu\psibar\ ,\\
(\psi\sigma^{\mu\nu}\theta)^\dagger & = & \thetabar\sigmabar^{\mu\nu}\psibar\ .
\end{eqnarray}

\paragraph{Derivatives:}

\begin{eqnarray}
\frac{\partial}{\partial\theta^\alpha} \theta^\beta 
& = & \delta_\alpha{}^\beta
\ ,\quad
\frac{\partial}{\partial\theta_\alpha} \theta_\beta 
= -\delta_\beta{}^\alpha
\ ,\\
\frac{\partial}{\partial\thetabar_\alphadot} \thetabar_\betadot 
& = & \delta^\alphadot{}_\betadot
\ ,\quad
\frac{\partial}{\partial\thetabar^\alphadot} \thetabar^\betadot 
= -\delta^\betadot{}_{\alphadot}
\ .
\end{eqnarray}

\section{BRS transformations}

On the
physical fields (i.e.\ fields carrying no ghost number) the BRS
transformations are the sum of gauge and supersymmetry transformations
and translations, where the transformation parameters have been
promoted to the ghost fields:
\begin{eqnarray}
\label{BRSStart}
sA{}_\mu & = & \partial_\mu c -ig [c,A_\mu]
             + i\epsilon\sigma_\mu\lambdabar
             -i \lambda\sigma_\mu\epsilonbar
\nonumber\\&&{}
             -i\omega^\nu\partial_\nu A_\mu
\ ,\\
s\lambda^\alpha & = & -ig \{c,\lambda^\alpha\}
             +\frac{i}{2} (\epsilon\sigma^{\rho\sigma})^\alpha
             F_{\rho\sigma} + i\epsilon^\alpha\, D
\nonumber\\&&{}
             -i\omega^\nu\partial_\nu  \lambda^\alpha
\ ,\\
s\lambdabar{}_\alphadot & = & -ig\{c,\lambdabar_\alphadot\}
             - \frac{i}{2} (\epsilonbar\sigmabar^{\rho\sigma})
             _\alphadot F_{\rho\sigma} + i\epsilonbar_\alphadot\, D 
\nonumber\\&&{}
             -i\omega^\nu\partial_\nu \lambdabar_\alphadot 
\ ,\\
\label{ChiralTransformationStart}
s\phi_i & = & -ig c\,\phi_i +\sqrt2\, \epsilon\psi_i - i\omega^\nu\partial_\nu \phi_i
\ ,\\
s\phi_i^\dagger & = & +ig (\phi^\dagger c)_i +\sqrt2\,
            \psibar_i\epsilonbar - i\omega^\nu\partial_\nu \phi_i^\dagger
\ ,\\
s\psi_i^\alpha & = & -ig c\,\psi_i^\alpha + \sqrt2\, 
         \epsilon^\alpha\, F_i
         -\sqrt2\, i (\epsilonbar\sigmabar^\mu)^\alpha D_\mu\phi_i 
\nonumber\\&&{}
         -i\omega^\nu\partial_\nu \psi_i^\alpha
\ ,\\
s{\psibar_i}_\alphadot & = & -ig ({\psibar}_\alphadot c)_i 
         - \sqrt2\,\epsilonbar_\alphadot\, F^\dagger_i 
         + \sqrt2\, i(\epsilon\sigma^\mu)_\alphadot (D_\mu\phi_i)^\dagger 
\nonumber\\&&{}
         -i\omega^\nu\partial_\nu {\psibar_i}_\alphadot
\label{ChiralTransformationEnd}
\ ,\\
sa & = & \sqrt2 \,\epsilon\chi -i\omega^\nu\partial_\nu a
\ ,\\
sa^\dagger & = & \sqrt2\,\chibar\epsilonbar -i\omega^\nu\partial_\nu
         a^\dagger
\ ,\\
s\chi^\alpha & = & \sqrt2\,\epsilon^\alpha \hat{f} - \sqrt2\,
         i(\epsilonbar\sigmabar^\mu)^\alpha \partial_\mu a
         -i\omega^\nu\partial_\nu \chi^\alpha
\ ,\\
s\chibar_\alphadot & = & -\sqrt2\,\epsilonbar_\alphadot
         \hat{f}^\dagger + \sqrt2\,i(\epsilon\sigma^\mu)_\alphadot
         \partial_\mu a^\dagger  -i\omega^\nu\partial_\nu
         \chibar_\alphadot
\ ,\\
sf & = & \sqrt2\, i\epsilonbar\sigmabar^\mu \partial_\mu \chi\
         -i\omega^\nu\partial_\nu f
\ ,\\
sf^\dagger & = & -\sqrt2\, i\partial_\mu \chibar\sigmabar^\mu\epsilon
         -i\omega^\nu\partial_\nu f^\dagger
\ .
\end{eqnarray}
Here we have used $A_\mu = T^a A_a{}_\mu$ and similar for $\lambda$,
$\lambdabar$, $F_{\rho\sigma}$, $D$, $c$, $\bar{c}$, $B$. Again, the
auxiliary fields $D$ and $F_i, F_i^\dagger$ are understood to be
eliminated by their equations of motion. 

The various (anti)commutation relations of the transformations are
encoded in the nilpotency equation
\begin{eqnarray}
s^2 & = & 0 + \mbox{field equations}
\label{Nilpotency}
\end{eqnarray}
if the BRS transformations of the ghosts are given by the structure
constants of the algebra and the ghosts have the opposite statistics
as required by the spin-statistics theorem \cite{BRS}:
\begin{eqnarray}
sc & = & -igc^2 + 2i\epsilon\sigma^\nu\epsilonbar A_\nu
         -i\omega^\nu\partial_\nu c
\ ,\\
s\epsilon^\alpha & = & 0
\ ,\\
s\epsilonbar^\alphadot & = &0
\ ,\\
s\omega^\nu & = & 2\epsilon\sigma^\nu\epsilonbar
\ .
\end{eqnarray}

The BRS transformations of the antighosts and $B$ fields read
\begin{eqnarray}
s\bar c_a & = & B_a - i\omega^\nu\partial_\nu \bar c_a
\ ,\\
sB_a & = & 2i\epsilon\sigma^\nu\epsilonbar \partial_\nu \bar c_a 
         -i\omega^\nu\partial_\nu B_a\ .
\label{BRSEnd}
\end{eqnarray}

\end{appendix}

\begin{flushleft}

\end{flushleft}
\end{document}